\documentclass{aa}  

\usepackage{graphicx}
\usepackage{txfonts}
\usepackage{xcolor}
\usepackage{ulem}
\usepackage{hyperref}
\usepackage{orcidlink}
\hypersetup{
  colorlinks,
  citecolor=blue,
  linkcolor=blue,
  urlcolor=blue}

\begin{document}

   \title{Violent mergers revisited: The origin of the fastest stars in the Galaxy}

   \author{R\"udiger Pakmor\inst{1}\thanks{rpakmor@mpa-garching.mpg.de}\orcidlink{0000-0003-3308-2420}
        \and
        Ken J. Shen\inst{2}\orcidlink{0000-0002-9632-6106}
        \and
        Aakash Bhat\inst{3}\orcidlink{0000-0002-4803-5902}
        \and
        Abinaya Swaruba Rajamuthukumar\inst{1}\orcidlink{0000-0002-1872-0124}
        \and
        Christine E. Collins\inst{4}\orcidlink{0000-0002-0313-7817}
        \and
        Cillian O'Donnell\inst{4}\orcidlink{0009-0007-5034-6420}
        \and
        Evan B. Bauer\inst{5,6}\orcidlink{0000-0002-4791-6724}
        \and
        Fionntan P. Callan\inst{7}\orcidlink{0000-0002-7975-8185}
        \and
        Friedrich K. R\"{o}pke\inst{8,9,10}\orcidlink{0000-0002-4460-0097}
        \and
        Joshua M. Pollin\inst{7}\orcidlink{0009-0005-6989-3198}
        \and
        Kate Maguire\inst{4}\orcidlink{0000-0002-9770-3508}
        \and
        Lindsey A. Kwok\inst{11}\orcidlink{0000-0003-3108-1328}
        \and
        Ravi Seth\inst{4}\orcidlink{0009-0002-9068-4006}
        \and
        Stefan Taubenberger\inst{12,1}\orcidlink{0000-0002-4265-1958}
        \and
        Stephen Justham\inst{1}\orcidlink{0000-0001-7969-1569}
        }

   \institute{Max-Planck-Institut f\"{u}r Astrophysik, 
              Karl-Schwarzschild-Str. 1, D-85748, Garching, Germany
         \and
             Department of Astronomy and Theoretical Astrophysics Center, University of California, Berkeley, CA 94720-3411, USA
         \and
             Institut f\"{u}r Physik und Astronomie, Universit\"{a}t Potsdam, Haus 28, Karl-Liebknecht-Str. 24/25, 14476 Potsdam-Golm, Germany
         \and
             School of Physics, Trinity College Dublin, College Green, Dublin 2, Ireland
         \and
             Lawrence Livermore National Laboratory, Livermore, California 94550, USA
         \and
             Center for Astrophysics | Harvard \& Smithsonian, 60 Garden Street, Cambridge, MA 02138, USA
         \and
             School of Mathematics and Physics, Queen's University Belfast, University Road, Belfast, BT7 1NN, UK
         \and
             Zentrum für Astronomie der Universität Heidelberg, Astronomisches Recheninstitut, M{\"o}nchhofstrasse 12--14, 69120 Heidelberg, Germany
         \and
             Heidelberger Institut f\"{u}r Theoretische Studien, Schloss-Wolfsbrunnenweg 35, 69118 Heidelberg, Germany
         \and
             Zentrum f\"{u}r Astronomie der Universit\"{a}t Heidelberg, Institut f\"{u}r Theoretische Astrophysik, Philosophenweg 12, 69120 Heidelberg, Germany
         \and
             Center for Interdisciplinary Exploration and Research in Astrophysics (CIERA), 1800 Sherman Ave., Evanston, IL 60201, USA
         \and
             TUM Department of Physics, Technical University Munich, Garching, Germany
             }

   \date{Received ; accepted}
 
   \abstract
   {
    Binary systems of two carbon-oxygen white dwarfs are one of the most promising candidates for the progenitor systems of Type Ia supernovae.
    Violent mergers, where the primary white dwarf ignites when the secondary white dwarf smashes onto it while being disrupted on its last orbit, were the first proposed double degenerate merger scenario that ignites dynamically.
    However, violent mergers likely contribute only a few per cent to the total Type Ia supernova rate and do not yield normal Type Ia supernova light curves.
    Here we revisit the scenario, simulating a violent merger with better methods, and in particular a more accurate treatment of the detonation.
    We find good agreement with previous simulations, with one critical difference. The secondary white dwarf, being disrupted and accelerated towards the primary white dwarf, and impacted by its explosion, does not fully burn. Its core survives as a bound object.
    The explosion leaves behind a $0.16\,\mathrm{M_\odot}$ star travelling $2800\,\mathrm{km/s}$, making it an excellent (and so far the only) candidate to explain the origin of the fastest observed hyper-velocity stars.
    We also show that before the explosion, $5\times10^{-3}\,\mathrm{M_\odot}$ of material consisting predominantly of helium, carbon, and oxygen has already been ejected at velocities above $1000\,\mathrm{km/s}$.
    Finally, we argue that if a violent merger made the hypervelocity stars D6-1 and D6-3, and violent mergers require the most massive primary white dwarfs in binaries of two carbon-oxygen white dwarfs, there has to be a much larger population of white dwarf mergers with slightly lower-mass primary white dwarfs. Because this population likely represents $\gg 10\%$ of the Type Ia supernovae rate, it can essentially only give rise to normal Type Ia supernovae.
   }


   \maketitle
   
   \nolinenumbers
%

\section{Introduction}
\label{sec:intro}

The progenitor systems and explosion mechanism of Type Ia supernovae are still debated. There is a general consensus only that they are thermonuclear explosions of carbon-oxygen white dwarfs in binary systems \citep{Liu2023,Ruiter2025}.

Mergers of two carbon-oxygen white dwarfs in a binary system have long been studied as one potential progenitor channel. However, the originally proposed `slow' merger channel in which the binary merges into a rotationally supported object that surpasses the Chandrasekhar mass and eventually contracts and explodes has been consistently shown not to work. It instead seems to lead to an accretion induced collapse to a neutron star \citep{Saio1985,Shen2012,Schwab2012,Schwab2021}.

Nevertheless, as first shown in \citet{Pakmor2010}, the binary system might already explode dynamically during the actual merger, as a "violent merger" explosion \citep{Pakmor2010,Pakmor2011,Pakmor2012,Sato2015,Sato2016}. In this scenario, when the secondary white dwarf is disrupted on its last orbit and accelerated towards the primary white dwarf, its impact on the surface of the primary white dwarf directly ignites a carbon detonation there. This detonation then sweeps through the primary white dwarf and the remains of the secondary white dwarf, and the explosion fully unbinds the system \citep{Pakmor2010,Pakmor2012,Sato2015}. Initially proposed as a channel to produce normal Type Ia supernovae, because it matches various early time observables reasonably well \citep{Pakmor2012}, violent mergers were soon shown to likely only produce peculiar explosions that fade more slowly and are more asymmetric than normal Type Ia supernovae \citep{Kromer2013b,Bulla2016}.

Shortly after the violent merger scenario was proposed, it was realised that the helium shell inherent to the surface of any carbon-oxygen white dwarf likely facilitates an explosion of at least the primary white dwarf of the binary via the double detonation mechanism long before the system actually merges \citep{Guillochon2010,Pakmor2013,Shen2014b}. This will likely work for most systems that have a massive enough helium shell on the primary white dwarf, and only fail for the most massive primary white dwarfs that have the smallest helium shells \citep{Shen2024}. Thus, the violent merger channel fell out of focus.

However, the violent merger scenario was later shown to explain many aspects of some peculiar Type Ia supernovae well. The over-luminous Type Ia supernova sub-class of `03fg-like' events have been connected to violent mergers where the extra luminosity compared to a normal Type Ia supernova is explained by interaction with the material of the disrupted secondary white dwarf \citep{Dimitriadis2022, Siebert2023, Siebert2024, Kwok2024}. They have also been suggested to explain the broad light curves and sub-luminous nature of the Type Ia supernova sub-class of `02es-like’ events \citep{Maguire2011, Ganeshalingam2012, Kromer2013b, Srivastav2023}. Both of these classes have been observationally shown to have flux excesses (`bumps') in their early light curves that are absent for other Type Ia supernova classes, pointing to a unifying explosion scenario \citep{Hoogendam2024}, which has been argued to be the violent merger model. 

Here, we revisit the violent merger scenario, simulating the explosion of a binary system consisting of two $1.1\,\mathrm{M_\odot}$ and $0.7\,\mathrm{M_\odot}$ carbon-oxygen white dwarfs with the moving mesh code \textsc{arepo} \citep{Arepo, Pakmor2016, Weinberger2020}. We are in particular able to follow the pre-merger evolution more faithfully and at better numerical resolution than previous simulations. Moreover, we model the nuclear burning in the detonation more accurately than the original simulations \citep{Pakmor2012b}. We summarise the methods and setup of our simulation in Section~\ref{sec:methods}.

We then show in Section~\ref{sec:explosion} that the inspiral, ignition, and explosion are in many aspects very similar to previous simulations of violent mergers. However, we find one critical difference: the secondary white dwarf is not fully burned. Despite being partially disrupted at the time of ignition and hit by the explosion, its core survives the explosion and remains bound. We characterise the bound remnant in Section~\ref{sec:remnant} and argue that this surviving, very low mass and fast moving star, is a good candidate to explain the extreme hyper-velocity stars D6-1 and D6-3 found with \textit{Gaia} \citep{Shen2018}.

Our new simulation models the initial inspiral for an unprecedented length of time. We exploit this to analyse the properties of the material that is expelled from the binary system during the last phase of the inspiral to form circumstellar material in Section~\ref{sec:csm}. We discuss the impact of our results for violent mergers, and in general for peculiar supernovae and normal Type Ia supernovae in Section~\ref{sec:discussion}. Finally, we conclude with a summary and an outlook in Section~\ref{sec:summary}.

\section{Methods}
\label{sec:methods}

Our simulation setup consists of three distinct steps. We first generate white dwarfs with the desired masses and a realistic composition profile by evolving helium-burning stars in isolation until they cease burning and cool as white dwarfs, using the stellar evolution code \textsc{mesa} \citep{Paxton2011}. Here, we follow the recipe used and described in detail in \citet{Shen2023}.

We then generate isolated 3D white dwarfs with the same radial profiles in the moving mesh magnetohydrodynamics code \textsc{arepo} \citep{Arepo,Pakmor2016,Weinberger2020} that we use to simulate the inspiral and explosion in a similar setup as other recent white dwarf merger simulations with \textsc{arepo} \citep{Pakmor2021,Pakmor2022,Burmester2023,Glanz2025}. \textsc{arepo} implements a second-order finite-volume scheme for magnetohydrodynamics on a moving Voronoi mesh. The mesh is evolved in an approximately Lagrangian way, that is, a set of mesh-generating points that move with the local gas velocity with small corrections to keep the mesh regular. Self-gravity is computed with the oct-tree method \citep{Springel2005}.

We employ explicit refinement and de-refinement of cells that deviate more than a factor of two from the target gas mass of $10^{-7}\,\mathrm{M_\odot}$. Additionally, we use a passive tracer to mark the helium shells on both white dwarfs to resolve them at a ten times better mass resolution of $10^{-8}\,\mathrm{M_\odot}$. This allows us to resolve even helium shells below $10^{-3}\,\mathrm{M_\odot}$.

We use the Helmholtz equation of state \citep{Timmes2000} that models an arbitrarily degenerate electron-positron gas, as well as ions, as fully ionised non-degenerate ideal gas with additional Coulomb corrections, and includes radiation. We also fully couple a $55$ isotope nuclear reaction network \citep{Pakmor2012,Pakmor2021} with JINA reaction rates \citep{Cyburt2010}. The nuclear reaction network is active for all cells with a temperature $T>10^6\,\mathrm{K}$. We use a standard burning limiter to disable nuclear burning in shocks to ensure that the unresolved detonation propagates at the correct speed \citep{Fryxell1989,Seitenzahl2009,Pakmor2021}.

To map the white dwarfs from \textsc{mesa} to \textsc{arepo} we first generate approximate 3D representations of the 1D density profiles of the white dwarfs with roughly round cells of approximately equal mass using HEALPix tessellations of spherical surfaces  \citep{Pakmor2012,Ohlmann2017}. We then map the precise density and composition profile of the white dwarfs, and evolve them for $10$ sound-crossing timescales. We relax them with a time-dependent friction term for the first $5$ sound-crossing times, and then evolve them freely for another $5$ sound-crossing times to make sure they are stable.

\section{Inspiral and explosion}
\label{sec:explosion}

\begin{figure}
    \centering
    \includegraphics[width=\linewidth]{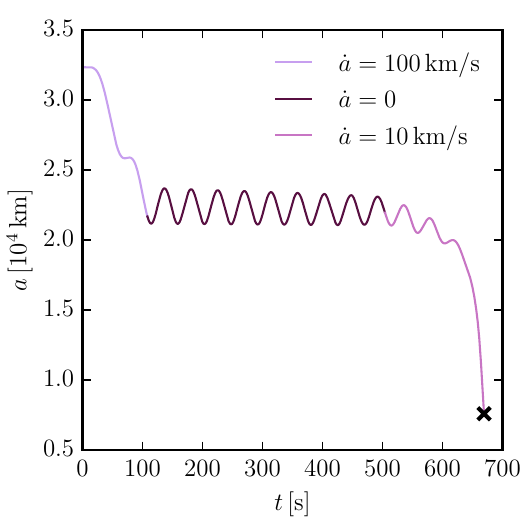}
    \caption{Time evolution of the separation between the two white dwarfs, computed as the distance between the centres of mass of the material originating from both white dwarfs. We first shrink the binary system at a constant rate of $100\,\mathrm{km/s}$ for $108\,\mathrm{s}$. We then evolve it for another $395\,\mathrm{s}$ self-consistently, that is, the total angular momentum in the simulation is conserved. In this phase mass transfer only shrinks it at a rate of $\approx 0.5\,\mathrm{km/s}$. We then actively shrink it again at a rate of $10\,\mathrm{km/s}$ until it merges $166\,\mathrm{s}$ later. The cross denotes the time of explosion.}
    \label{fig:inspiral}
\end{figure}

\begin{figure*}
    \centering
    \includegraphics[width=\textwidth]{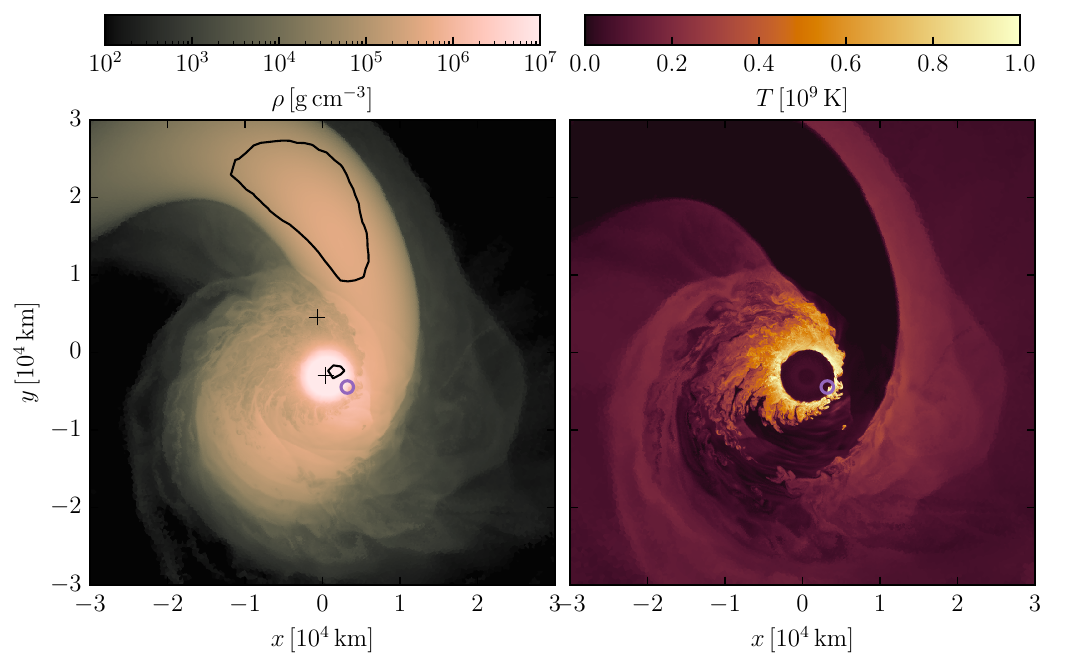}
    \caption{Density (left panel) and temperature (right panel) slices of the binary system during the merger at the time when it reaches conditions for carbon ignition. The purple circle shows the point where $32$ cells reach a temperature of $10^9\,\mathrm{K}$ at a density of $2\times10^6\,\mathrm{g\,cm^{-3}}$. We ignite a detonation there (see text). The contours in the left panel show the origin of $95\%$ of the material that will become the bound remnant after the explosion. The crosses in the left panel show the centres of mass of the material originating from each white dwarf.}
    \label{fig:ignition}
\end{figure*}

Our initial binary system consists of a primary white dwarf with a total mass of $1.1\,\mathrm{M_\odot}$ and a secondary white dwarf with a total mass of $0.7\,\mathrm{M_\odot}$. The primary white dwarf comes out to consist of essentially equal amounts of $0.54\,\mathrm{M_\odot}$ each of carbon and oxygen, mixed with $0.016\,\mathrm{M_\odot}$ of $^{22}\mathrm{Ne}$, and a helium shell of $3\times10^{-4}\,\mathrm{M_\odot}$. It has a radius of $4600\,\mathrm{km}$, and its helium shell, defined as range in which the helium mass fraction is larger than $0.1$, has a width of $200\,\mathrm{km}$. The secondary white dwarf is made of $0.32\,\mathrm{M_\odot}$ of carbon, $0.36\,\mathrm{M_\odot}$ of oxygen, $0.011\,\mathrm{M_\odot}$ of $^{22}\mathrm{Ne}$, and a helium shell of $0.009\,\mathrm{M_\odot}$. It has a radius of $7600\,\mathrm{km}$ and a helium shell with a width of $1100\,\mathrm{km}$.

After having relaxed both white dwarfs separately, we put them together in a binary system in co-rotation with an initial separation of $3.2\times10^9\,\mathrm{cm}$, equivalent to an initial orbital period of $74\,\mathrm{s}$. At this distance, the secondary white dwarf does not fill its Roche lobe. The inspiral then proceeds in three steps. We show the full evolution of the binary separation with time in Figure~\ref{fig:inspiral}.

In the first phase, we apply an artificial gravitational wave-like source term that removes angular momentum and shrinks the binary separation at a constant rate \citep{Pakmor2021,Pakmor2022}. In this phase we shrink the binary system at a rate of $\dot{a}=100\,\mathrm{km/s}$ for $108\,\mathrm{s}$, until it reaches a separation of $2.2\times10^9\,\mathrm{cm}$, equivalent to an orbital period of $41\,\mathrm{s}$. At this time the density at the inner Lagrange point L1 reaches $10^4\,\mathrm{g\,cm^{-3}}$ (which is typically sufficient to quickly lead to helium ignition on the primary white dwarf if there is sufficient helium on it; see for example \citealt{Pakmor2022}). Note that the small eccentricity during this phase is a result of the relatively fast initial inspiral. To avoid this, we likely have to ensure that the distance only changes by a small amount in every orbit, that is $\dot{a} \ll a / P \approx 500\,\mathrm{km/s}$, where $P$ is the orbital period. However, we currently cannot afford the computational cost to evolve the system this slowly all the way from the time the secondary white dwarf fills its Roche-lobe to the explosion.

In the second phase, we then continue to evolve the binary system for $395\,\mathrm{s}$ without the artificial angular momentum loss term, that is, we conserve the total angular momentum of the system. In this phase, the binary separation continues to shrink as a result of mass transfer, but only at a very small rate of $\dot{a}\lesssim 0.5\,\mathrm{km/s}$, barely visible in Figure~\ref{fig:inspiral}.

Realising that computational limitations will not allow us to evolve the system to the point where it merges this way, we enable the angular momentum loss term again in the third phase. In this phase, we shrink the system again actively at a constant rate, but reduce it to a tenth of the initial rate, that is, to $\dot{a}=10\,\mathrm{km/s}$, which is still $20$ times faster than the physical rate. We keep this term active until the system merges and plausibly ignites $166\,\mathrm{s}$ later. Having the angular momentum loss term active until the system explodes might affect the detailed properties of the material ejected prior to the merger and of the secondary white dwarf during the merger. This is mostly limited by computational constraints and should be investigated in more detail in the future.

At the time of ignition, the total angular momentum of the binary system, that is the total angular momentum of all material in our simulation relative to the centre and in the rest frame of the binary system, has decreased to $6.3\times10^{50}\,\mathrm{g\,cm^2\,s^{-1}}$ from $6.5\times10^{50}\,\mathrm{g\,cm^2\,s^{-1}}$ during the intermediate phase that conserved angular momentum. During the second and third inspiral phase, all helium initially present on the surface of the secondary white dwarf is transferred to the primary white dwarf. $5\times10^{-3}\,M_\odot$ of this helium are burned in bursts on the surface of the primary white dwarf. This ejects the ashes of the helium burning as well as the remaining unburned helium from the system (see also Section~\ref{sec:csm}).
In contrast to previous simulations with more massive helium shells on the primary white dwarf \citep[for example][]{Pakmor2013,Pakmor2022}, the helium shell on our $1.1\,\mathrm{M_\odot}$ primary white dwarf is too small to sustain a helium detonation \citep{Shen2024}. Therefore, there is also no core ignition via the double detonation scenario.

At $668\,\mathrm{s}$, the secondary white dwarf is being disrupted and accelerated towards the primary. As it hits the primary white dwarf directly, material on the surface of the primary white dwarf is compressed and heated up to reach conditions favourable for carbon ignition, similar to previous violent merger simulations \citep{Pakmor2010,Pakmor2012,Sato2015}. Figure~\ref{fig:ignition} shows slices of density and temperature of the binary system in the plane of rotation at the time of ignition. The centres of mass of the material originating from both white dwarfs are denoted by black crosses. The centre of mass of the material originating from the secondary white dwarf directly show that it is already quite advanced in the process of being disrupted, as its centre is significantly offset from where the density peak in its tail. The purple circle shows the spot that contains $32$ cells with a density larger than $\rho=2\times10^6\,\mathrm{g\,cm^{-3}}$ as well as a temperature above $T=10^9\,\mathrm{K}$. It provides plausible conditions to ignite a carbon detonation \citep{Seitenzahl2009}.

\begin{figure*}
    \centering
    \includegraphics[width=0.97\textwidth]{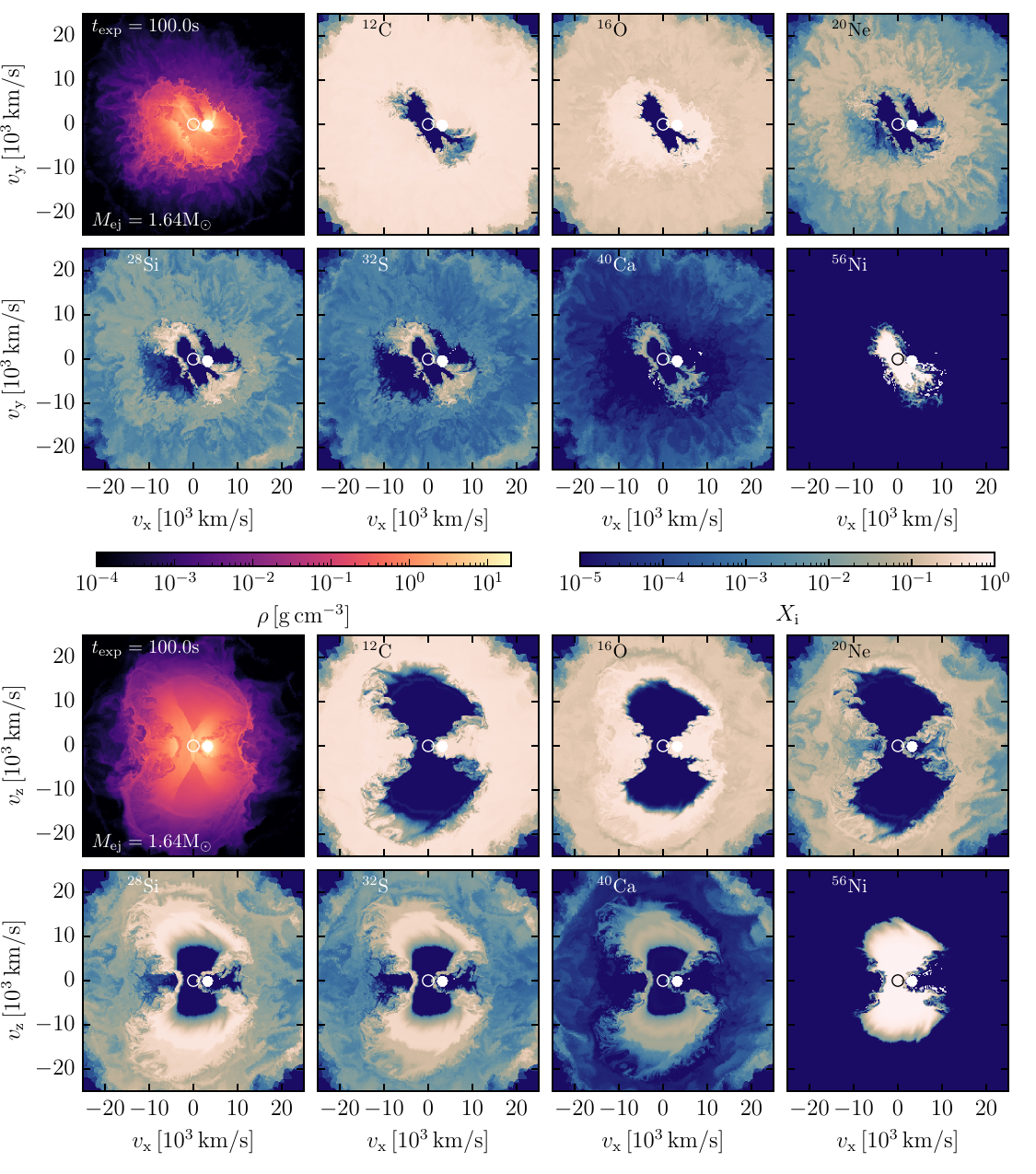}
    \caption{Slices through the ejecta $100\,\mathrm{s}$ after the explosion in the plane of rotation of the binary system (top rows) and perpendicular to it (bottom rows). At this time the ejecta are fully in homologous expansion. The panels show the density in the top left panel and mass fractions of different isotopes in the other panels. The open circle indicates the centre of the ejecta. The filled white circle appears where the material of the bound remnant is located, which has been removed here. The structures look very similar to older violent merger simulations, showing a clear global asymmetry and material originating from the secondary white dwarf close to the centre.}
    \label{fig:ejecta}
\end{figure*}

The ignition is not resolved by many orders of magnitude in spatial scale, so we put in a detonation at this spot by heating all material within a sphere of radius $100\,\mathrm{km}$ to a temperature of $5\times10^9\,\mathrm{K}$. This injects a total energy of $4.8\times10^{46}\,\mathrm{erg}$ into $188$ cells with a total mass of $5\times10^{-6}\,\mathrm{M_\odot}$ and starts a detonation. The detonation burns through the primary white dwarf and releases $1.7\times10^{51}\,\mathrm{erg}$ of nuclear energy. The explosion destroys the primary white dwarf. Its ejecta impact the secondary white dwarf that is already in a state of being disrupted, and unbind most of it as well. In contrast to previous violent merger simulations \citep{Pakmor2010,Pakmor2012}, the secondary white dwarf is not fully burned. Rather, a small fraction ($0.16\,M_\mathrm{\odot}$) of its core remains bound and even accretes a small amount of the ashes of the primary white dwarf on its surface. The location of the material that will make up the bound remnant is indicated with black contours in the left panel of Figure~\ref{fig:ignition}. Almost all of the material that will make up the bound remnant is contained in the upper contour that only contains material that originates from the secondary white dwarf. The lower contour that marks material which originates from the primary white dwarf only contributes $0.002\,\mathrm{M_\odot}$ to the bound remnant, or roughly $1\%$ of its mass. Thus, only a tiny fraction of the mass covered by the lower contour will become part of the bound remnant. We will discuss the properties of this remnant in detail in Section~\ref{sec:remnant}.

The main difference to previous violent merger models of a similar system \citep{Pakmor2012} is that they used the level-set method to model the detonation. That method models the detonation as a shock discontinuity that propagates with a given velocity and instantaneously burns material that is crossed by the detonation. It assumes that the composition of the material after it has been burned only depends on the pre-shock density, which is tabulated. However, this approximation ignores that the burning also depends on the strength of the shock, which can change as the shock propagates. A stronger shock compresses the material more before it is burned, and the nuclear burning proceeds further. The nuclear burning with the level-set method is typically calibrated to a centrally ignited spherical white dwarf. In this setup the shock travels down the density gradient and becomes stronger with time. Critically, a specific shock strength at each density is implicitly baked into this calibration. Because the detonation starts at the densest point, this setup for calibration likely leads to the strongest possible detonation shock, and therefore the furthest nuclear burning.

In our simulation, in contrast, the detonation is directly modelled by solving the full time-dependent reactive Euler equations. That is, in every time step a nuclear reaction network is fully coupled to the equations of hydrodynamics. The composition of each cells changes following the time integration of the nuclear reaction rates, and the nuclear energy release directly changes the internal energy of the cell. 

Despite not resolving the physical width of the carbon detonation by many orders of magnitude, this method in principle reproduces the change of composition and energy release from nuclear burning from the detonation on the scales we resolve \citep{Fryxell1989, Seitenzahl2009, Pakmor2013, Kushnir2020b}. One caveat is that because we do not resolve the physical width of the detonation shock, we need to limit nuclear burning in the shock itself \citep{Fryxell1989}. Here, we configure the limiter to label pressure jumps relatively aggressively as shocks \citep{Pakmor2021}, which usually leads to the furthest burning and largest energy release compared to milder limiters.

In our simulation the explosion of the primary white dwarfs launches a strong shock towards the secondary white dwarf. This shock is strongest when it just hits the secondary white dwarf that is already being disrupted. At this time, the central density of the secondary white dwarf has already dropped below $5\times10^5\,\mathrm{g\,cm^{-3}}$. The shock burns some of the carbon in the outer layers of the secondary white dwarf to neon and heavier intermediate mass elements. However, the shock weakens as it travels up the density gradient and the energy release from nuclear burning is insufficient to sustain the strength of the shock at these densities \citep{Dunkley2013}. Nuclear burning then essentially ceases in the core of the secondary white dwarf, and a fraction of the core remains bound and unburned. Note that because of our aggressive burning limiter, it is unlikely that a different treatment of the unresolved detonation in \textsc{arepo} would burn the whole secondary white dwarf.

\begin{figure*}
    \centering
    \includegraphics[width=\textwidth]{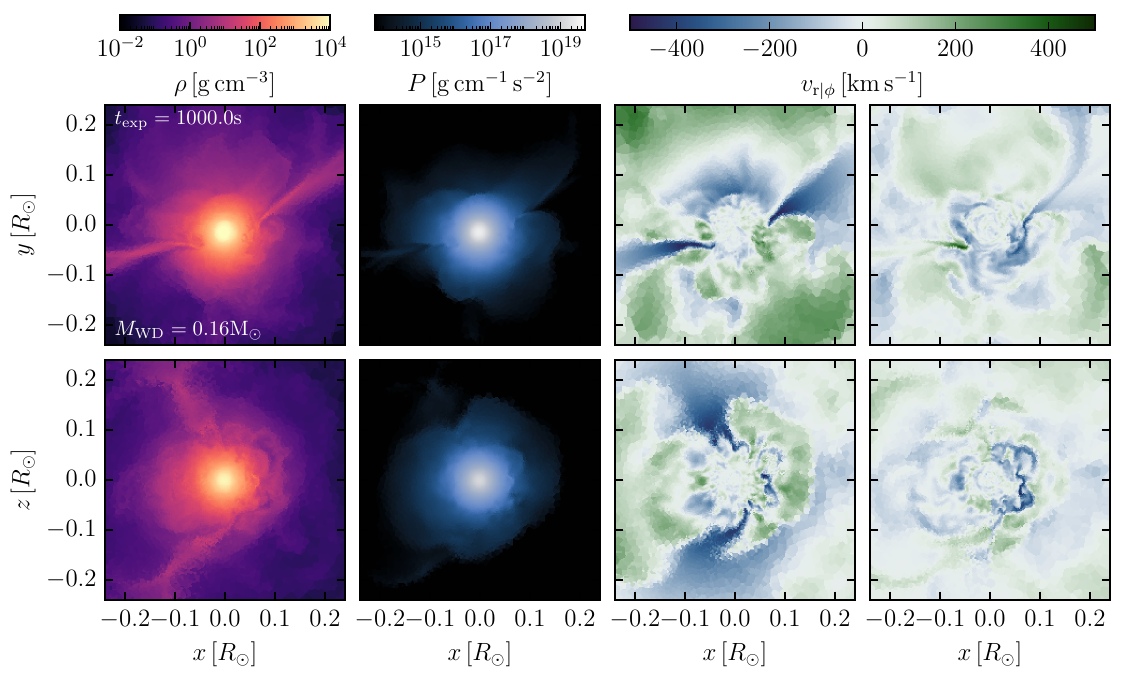}
    \caption{Slices of density (first column), pressure (second column), radial velocity (third column), and azimuthal velocity (fourth panel) of the bound remnant $1000\,\mathrm{s}$ after the explosion. The top row shows slices in the plane of rotation of the initial binary system, the bottom row slices perpendicular to it. The velocities are relative to the rest frame velocity of the bound remnant. The inner part of the $0.16\,\mathrm{M_\odot}$ bound remnant is essentially spherical, the outer parts are clearly not.}
    \label{fig:remnant}
\end{figure*}

Similar to previous violent merger simulations the ejecta of the primary expand much more easily in the vertical direction compared to the plane of rotation, where they encounter the material of the partially disrupted secondary white dwarf. This also implies that the impact of the ejecta of the primary white dwarf on the secondary white dwarf is weaker compared to isotropic expansion of the ejecta of the primary white dwarf. The material that is unbound from the secondary from the impact and partial burning is then incorporated into the central parts of the ejecta of the primary white dwarf. The density structure of the ejecta and its composition in homologous expansion $100\,\mathrm{s}$ after ignition are shown in Figure~\ref{fig:ejecta}. The ejecta contain a total mass of $1.64\,\mathrm{M_\odot}$, and notably $0.72\,M_\odot$ of radioactive $^{56}\mathrm{Ni}$, $0.03\,M_\odot$ of stable $^{58}\mathrm{Ni}$, $0.02\,M_\odot$ of $^{40}\mathrm{Ca}$, $0.16\,M_\odot$ of $^{28}\mathrm{Si}$, $0.29\,M_\odot$ of $^{16}\mathrm{O}$, $0.03\,M_\odot$ of $^{20}\mathrm{Ne}$, and $0.19\,M_\odot$ of $^{12}\mathrm{C}$. Most of the material of the initial binary system is part of the ejecta. The total centre-of-mass velocity of the ejecta with respect to the rest frame of the original binary system is only $290\,\mathrm{km/s}$.

The ejecta composition and structure is very similar to previous violent merger simulations \citep{Pakmor2012,Sato2015}. Most of the ejecta material originating from the secondary white dwarf is unburned. However, some of its carbon has been burned to neon, which, together with unburned carbon and oxygen, is located close to the very centre of the ejecta. In particular, the burning of the secondary white dwarf still produces $0.025\,\mathrm{M_\odot}$ of $^{20}\mathrm{Ne}$, about a third of what was produced in the old simulation in which the secondary was completely burned \citep{Pakmor2012}. Neon in the centre of the ejecta is of particular interest because it has so far only been seen in JWST observations of nebular spectra of SN~2022pul, a 03fg-like Type Ia supernovae that has been argued to be a violent merger \citep{Blondin2023,Kwok2024}. Moreover, SN~2022pul has also been shown to have carbon in the centre of the ejecta \citep{Liu2025}, that is naturally explained by our ejecta structure. However, the detailed morphology of the central ejecta is complicated, so only future 3D radiative transfer simulations of the nebular phase \citep{Pollin2025} of this model will be able to show if our model produces similar neon and carbon lines in the nebular spectrum as observed for SN~2022pul.

Above and below the orbital plane, the ashes of the primary white dwarf spread much further out in velocity space than in the plane of rotation. The iron group elements reach velocities above $10^4\,\mathrm{km/s}$ in the vertical direction. We leave detailed radiation transfer post-processing and a comparison of synthetic observables with observations and previous simulations to future work. We expect synthetic observables of our explosion to show significant line of sight variation and overall polarisation, similar to previous violent merger models \citep{Pakmor2012,Bulla2016}.

\begin{figure*}
    \centering
    \includegraphics[width=\textwidth]{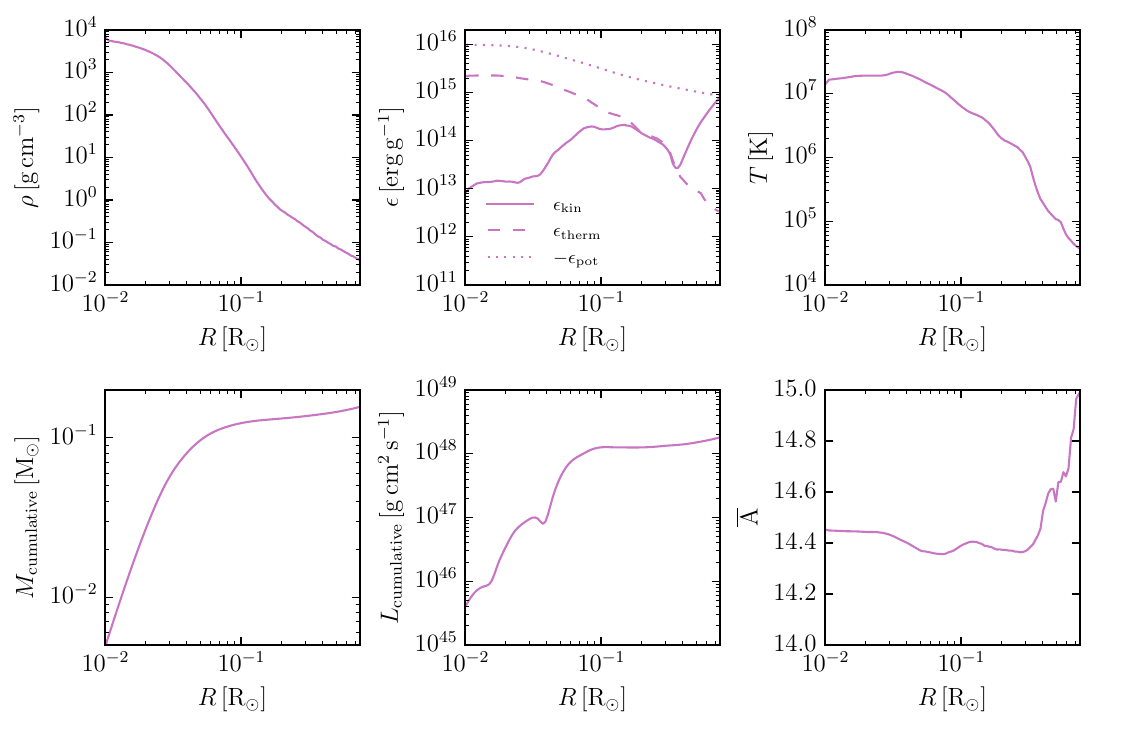}
    \caption{Radial profiles of the bound remnant $1000\,\mathrm{s}$ after the explosion. The panels show from left to right and top to bottom profiles of density, specific energies, temperature, cumulative mass, cumulative angular momentum, and mean atomic weight. The essentially spherical part of the remnant seen in Figure~\ref{fig:remnant} extends out to a radius of about $0.04\,\mathrm{R_\odot}$. It is essentially non-rotating and contains about half of the bound mass of the remnant. The outer parts are rotating, cold, and contain a little bit of the heavier elements from the ashes of the explosion.}
    \label{fig:remnant_profile}
\end{figure*}

\section{Remnant}
\label{sec:remnant}

The main difference between our simulation and previous violent merger simulations is the presence of a surviving bound remnant. The remnant is the remainder of the core of the secondary white dwarf that is not disrupted from the impact of the explosion (see Figure~\ref{fig:ignition}). When the secondary white dwarf is hit by the explosion of the primary white dwarf, it is already in the process of being disrupted and accelerated towards the primary white dwarf. The material that remains bound is therefore significantly faster than the orbital velocity of the binary at the time of ignition and moves with a velocity of $2800\,\mathrm{km/s}$ relative to the rest frame of the original binary system.

The high velocity (${\gg}{2000}\,\mathrm{km/s}$) of the surviving star is so far a unique feature of the violent merger scenario. It arises from an initial orbital velocity of $\sim2000\,\mathrm{km/s}$ owing to the high total mass of the binary system \citep{Shen2018}, and the additional acceleration of the secondary white dwarf towards the primary white dwarf on its last orbit. In contrast, a massive binary system in which the primary white dwarf explodes via a double detonation explosion before the secondary is disrupted \citep{Pakmor2013,Shen2018,Braudo2024} has an upper limit on the ejection velocity of the surviving star of $\approx2000\,\mathrm{km/s}$. A similar maximum ejection velocity limit applies for an explosion of the primary white dwarf when the secondary is disrupted in a binary system with a much lower total mass \citep{Glanz2025}.

The high velocity of our surviving star can explain the velocities of D6-1 and D6-3, the fastest of the objects found in the original search for hyper-velocity stars \citep{Shen2018}, as well as the fastest hyper-velocity stars found since then \citep{ElBadry2023}. Moreover, more detailed modelling indicates masses for D6-1 and D6-3 of only $0.1\,\mathrm{M_\odot}$-$0.2\,\mathrm{M_\odot}$ \citep{Shen2025}, which are also fully consistent with our scenario. These low masses, however, are inconsistent with double detonation explosion scenarios where the secondary white dwarf is unscathed \citep{Bauer2021,Pakmor2022} or only slightly distorted and stripped \citep{Glanz2025}. In those scenarios the surviving star still has a mass of $0.5\,\mathrm{M_\odot}$ or more. We will discuss the potential of the violent merger scenario to produce other very fast but likely more massive hyper-velocity stars \citep{ElBadry2023} at the end of this Section.

We show slices of the bound material $1000\,\mathrm{s}$ after the explosion in Figure~\ref{fig:remnant}. At this time, the inner part of the bound remnant out to a radius of ${\approx}\,0.05\,\mathrm{R_\odot}$ has relaxed to an essentially spherically symmetric state and does not rotate. The outer parts at larger radii are still far from spherical and feature significant radial and azimuthal velocity components. However, due to computational constraints, we cannot evolve it in \textsc{arepo} until it would also be relaxed in the outer parts.

We show radial profiles of the bound remnant at $1000\,\mathrm{s}$ after the explosion in Figure~\ref{fig:remnant_profile}. At this time, $0.16\,\mathrm{M_\odot}$ are bound. Roughly $0.1\,\mathrm{M_\odot}$ of them are contained within the inner $0.05\,\mathrm{R_\odot}$ (see lower left panel of Figure~\ref{fig:remnant_profile}) where they have already relaxed to be spherically symmetric (see Figure~\ref{fig:remnant}). The additional bound mass is spread out to much larger radii, and the total bound mass is only reached at a solar radius. At these larger radii the material is significantly supported by rotation and radial motions (see upper middle panel of Figure~\ref{fig:remnant_profile}). The outermost parts appear technically unbound, though this is an artifact of the spherical averaging of the remnant that is far from spherical symmetric at these large radii. The outer parts also contain almost all of the angular momentum of the bound remnant (see lower middle panel of Figure~\ref{fig:remnant_profile}). As shown in the profile of the mean atomic weight (see lower right panel of Figure~\ref{fig:remnant_profile}) the ejecta are for the most part just made from the unburned material of the secondary white dwarf. The bound remnant consists of $0.07\,\mathrm{M_\odot}$ of carbon and $0.09\,\mathrm{M_\odot}$ of oxygen. 
Only the very outermost layers contain some of the burned ashes of the primary white dwarf, including $10^{-4}\,\mathrm{M_\odot}$ of $^{56}\mathrm{Ni}$. The bound remnant features temperatures of $\approx10^7\,\mathrm{K}$ for most of its relaxed parts (see upper right panel of Figure~\ref{fig:remnant_profile}).

While the high velocity and low mass of the bound remnant make it a prime candidate to explain the hyper-velocity stars D6-1 and D6-3 \citep{Shen2018,Shen2025}, detailed modelling is required to fully confirm or reject this association \citep{Bhat2025,Glanz2025}. This modelling will be complicated by the significant deviation from spherical symmetry of a third of the material of the bound remnant even $1000\,\mathrm{s}$ after the explosion. Notably, the surviving star in the violent merger scenario might have been detected already in a SN~2020hvf, which is part of the subclass of 03fg-like ``super-Chandrasekhar'' Type Ia supernovae \citep{Siebert2023}.

It is hard to assess how robust the results are with respect to the mass of the bound remnant. It will certainly depend on the detailed timing of ignition. Significantly later ignition would likely result in no bound remnant, because the secondary white dwarf would have been fully disrupted already. Significantly earlier ignition would probably leave a more massive remnant. Significantly earlier ignition seems unlikely, because the temperature increases quickly before the time when we ignite. Significantly later ignition does not seem to be likely either, because the temperature does not increase significantly beyond the temperature of $10^9\,\mathrm{K}$ we chose for ignition for the next $5\,\mathrm{s}$.

The choice of burning limiter will probably not have a significant effect on the mass of the bound remnant. The limiter is most important for detonations at high densities, but will make almost no difference at the low densities at which the secondary white dwarf is partially burned. The main path to more massive bound remnant is probably to start from a binary system with a more massive secondary white dwarf. This makes it even more rare though.

\section{Circumstellar material}
\label{sec:csm}

\begin{figure*}
    \centering
    \includegraphics[width=\textwidth]{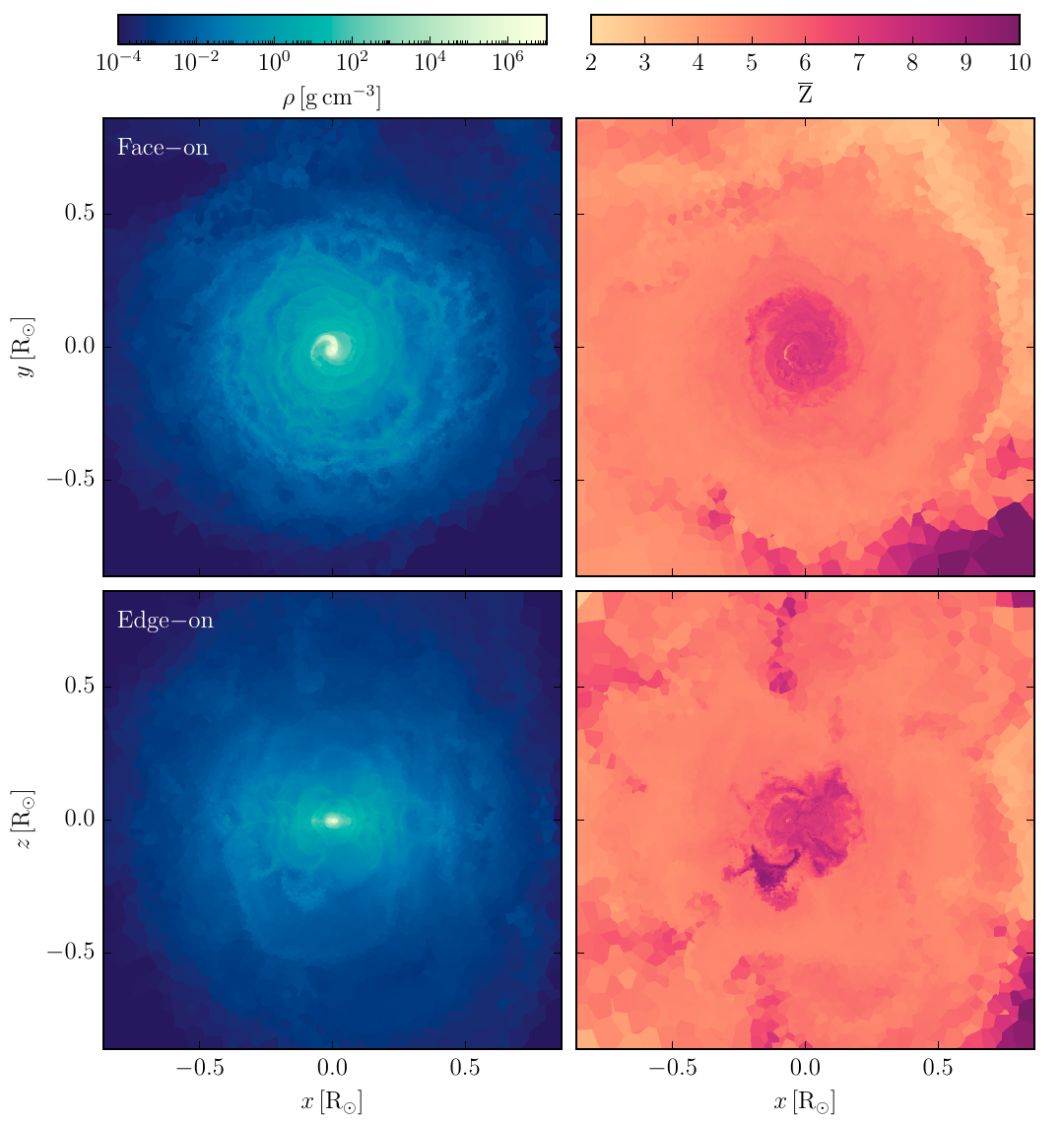}
    \caption{Slices of density (left column) and mean charge (right column) of the whole simulation box at the time of explosion. The top row shows slices in the plane of rotation of the binary system, the bottom row shows slices perpendicular to it. There clear outflows to large distances from the binary system even before the explosion. These outflows contain some heavier metals at various distances as a result of explosive helium burning on the surface of the primary white dwarf during the inspiral.}
    \label{fig:environment}
\end{figure*}

\begin{figure}
    \centering
    \includegraphics[width=\linewidth]{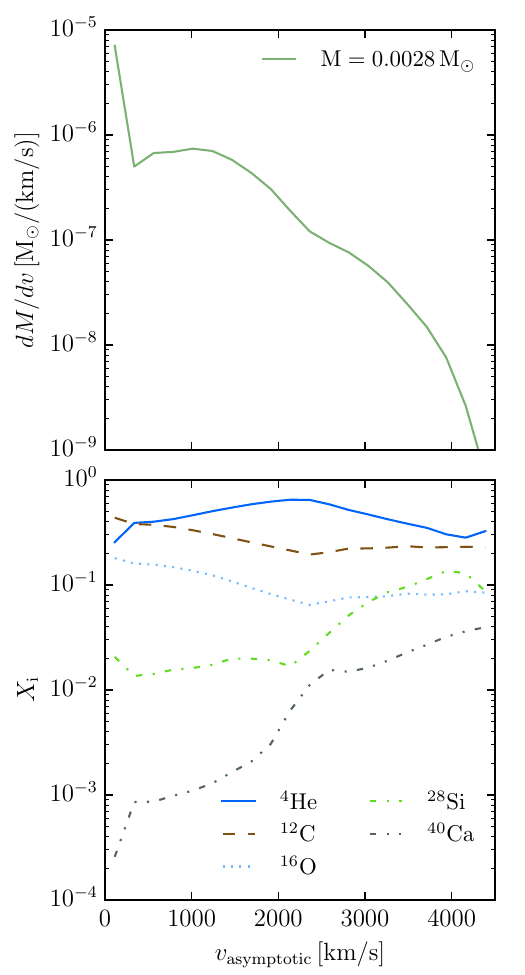}
    \caption{Mass (top panel) and abundance (bottom panel) profiles in velocity space of the unbound material at the time of the explosion. The ejected mass has a typical velocity of $1500\,\mathrm{km/s}$ and consists mostly of helium, carbon, and some intermediate mass elements at the largest velocities.}
    \label{fig:preexplosion}
\end{figure}

\begin{figure}
    \centering
    \includegraphics[width=\linewidth]{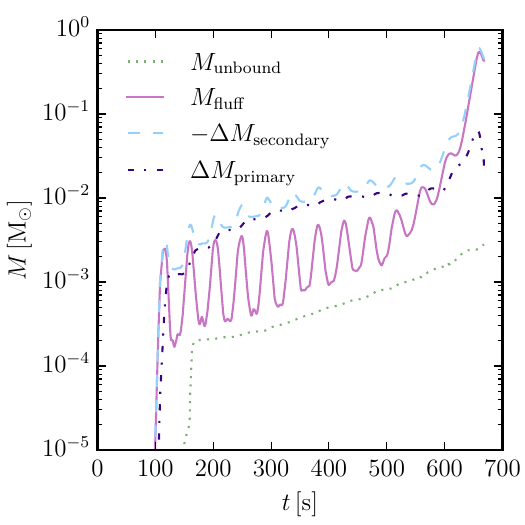}
    \caption{Time evolution of mass transfer. We separate all mass the material lost from the secondary white dwarf ($\Delta M_\mathrm{secondary}$), the mass gained by the primary white dwarf ($\Delta M_\mathrm{primary}$), the mass only associated with the binary system as a whole ($M_\mathrm{fluff}$), and the unbound material $M_\mathrm{unbound}$. Before the explosion, about $3\times10^{-3}\,\mathrm{M_\odot}$ have become unbound already.}
    \label{fig:masstransfer}
\end{figure}

The very high mass resolution of our simulation, in particular in the outer layers of the secondary white dwarf of $10^{-8}\,\mathrm{M_\odot}$, and the long inspiral phase prior to the explosion of more than $10$ orbits allow us to have a look at the material that is ejected from the system during mass transfer before the actual explosion. Note that this phase is still significantly accelerated in our simulation by the artificial angular momentum loss term. Even in the third phase, when the additional term reduces the separation at a rate of $10\,\mathrm{km/s}$, this still shrinks the binary system at least $20$ times faster than in the phase before via mass transfer only without the additional term. So our results are most likely a lower limit on the pre-explosion mass loss.

For a visual impression of the circumstellar material, we show slices of density and mean atomic weight in the plane of rotation and perpendicular to it out to a distance of $0.8\,\mathrm{R_\odot}$ at the time of ignition in Figure~\ref{fig:environment}. We clearly see, that material has been transported out to large distances (for reference, the separation of the binary system at this time is roughly $10^{-2}\,\mathrm{R_\odot}$). The circumstellar material is slightly denser in the plane of rotation but extends to similarly large radii in all directions. It is dominated by unburned helium, carbon and oxygen. Its inner parts also contain heavier elements that originate from helium burning on the surface of the primary white dwarf. This material was clearly ejected later than the outermost material, so there was significant mass loss already during the accretion phase prior to any helium burning.

We quantify the properties of the circumstellar material that is unbound from the binary system at the time of ignition in Figure~\ref{fig:preexplosion}. The top panel shows a histogram of mass in velocity space. Here, the velocity is the asymptotic velocity at infinity computed from the current velocity of each unbound cell and its local gravitational potential. The total mass of the unbound material is $3\times10^{-3}\,\mathrm{M_\odot}$, most of it at velocities below $2000\,\mathrm{km/s}$ with a tail to larger velocities up to $5000\,\mathrm{km/s}$. This material is dominated by helium with significant mass fractions of carbon, and a smaller amount of oxygen, representing the composition of the outer layers of the secondary white dwarf. It has only minor contributions of heavier elements. The material mostly originates from the unmodified material of the outer layers of the secondary white dwarf. The material at the largest velocities has a significantly higher mass fraction of silicon and calcium; products of helium burning on the surface of the primary white dwarf. The higher velocities are a consequence of the additional extra energy release from nuclear burning, though the total mass at these large velocities is tiny.

We show the time evolution of mass transfer and mass loss from the binary system in Figure~\ref{fig:masstransfer}. We first compute which material is unbound from the binary system, that is the mass of all cells with a positive total energy $\epsilon_\mathrm{tot}=\epsilon_\mathrm{thermal}+\epsilon_\mathrm{kinetic}+\epsilon_\mathrm{potential}$. The remaining material is bound to the binary system. We further subdivide the bound material by computing the potential at the inner Lagrange point L1. We then count all material at a higher potential as ``fluff'' that is only bound to the binary system as a whole, but cannot easily be associated with one of the individual white dwarfs. We then assign all bound material at a deeper potential than the potential at L1 to either white dwarf, depending on which side of L1 it is.

Figure~\ref{fig:masstransfer} shows that the secondary white dwarf is losing mass at a rate of order $10^{-5}\,\mathrm{M_\odot/s}$ in our simulation. Most of the material becomes bound to the primary white dwarf and increases its mass. About $10\%$ of the material lost from the secondary white dwarf before its last orbit becomes fluff, bound to the binary system but not obviously to one of the individual white dwarfs. Because the separation of the binary system oscillates slightly (see also Figure~\ref{fig:inspiral}), the potential at L1 oscillates as well. This leads to oscillations in the estimates of the mass already lost from the secondary white dwarf and the mass associated with the fluff.

A few per cent of the material that is lost from the secondary white dwarf becomes unbound from the binary system. At the time of ignition, this accumulates to $3\times10^{-3}\,\mathrm{M_\odot}$. This material, in particular if it reaches significant distances from the binary system when it explodes, might significantly affect observables of the supernova. We look at this as a first step to understand how much material is actually ejected during the inspiral phase and what properties it might have. Again, because of the accelerated inspiral, this is likely a lower limit on the amount of mass lost, and certainly a lower limit on the distance at which we would find this material at the time of explosion. When we compute the unbound mass ignoring its thermal energy, that is only comparing kinetic and potential energy, the unbound mass evolves similarly to Figure~\ref{fig:masstransfer}, but is about a factor of two smaller.

The ejecta velocities and total mass are qualitatively consistent with lower resolution smoothed particle hydrodynamics simulations of mergers of two carbon-oxygen white dwarfs \citep{Raskin2013}. In our simulation the material is ejected over a few hundred seconds with a typical velocity of  $1000\,\mathrm{km/s}$. Since the inspiral in our simulation is faster by at least a factor of $20$ than in reality (see Section~\ref{sec:explosion}), the inspiral phase should last correspondingly longer. The ejected material can thus reach distances of $10^{12}\,\mathrm{cm}$ as a conservative estimate. The ejecta of the explosion, that are roughly ten times faster than the material ejected during inspiral, will reach the circumstellar material likely on the first day after explosion. The interaction between the ejecta and this material might produce a bump in the light curve at early times. The amount and properties of the circumstellar material in our simulation are roughly in the regime of the circumstellar material argued to be around the two 03fg-like Type Ia supernovae SN~2021zny ($0.04\,\mathrm{M_\odot}$ at a distance of $10^{12}\,\mathrm{cm}$, \citealt{Dimitriadis2023}) and SN~2022ilv ($10^{-3}$ to $10^{-2}\,\mathrm{M_\odot}$ at a distance of $10^{13}\,\mathrm{cm}$, \citealt{Srivastav2023}). In the future we will need better modelling of the ejection of material during inspiral as well as detailed modelling of the interaction of the ejecta with the previously ejected material to understand these early-time observations.

\section{Discussion}
\label{sec:discussion}

The critical condition for a binary system of two sufficiently massive carbon-oxygen white dwarfs to explode via the violent merger scenario is that the primary white dwarf does not explode before already via the double detonation mechanism. If a double detonation ignition of the primary white dwarf is possible, it will always happen earlier, because a helium detonation ignites at much lower temperatures and densities than a carbon detonation.

In the simulation we present here a double detonation of the primary white dwarf is absent because its surface helium layer is too thin to support a helium detonation. This is the natural outcome for carbon-oxygen white dwarfs with a mass above $\gtrsim 1.0\,\mathrm{M_\odot}$ that did not accrete additional helium after formation \citep{Shen2024}. This selection criterion on massive primary white dwarfs helps to explain the brightness of 03fg-like (``super-Chandrasekhar'') objects, that have been associated with violent mergers \citep{Taubenberger2013,Taubenberger2013b,Taubenberger2017,Dimitriadis2022,Srivastav2023,Siebert2023,Siebert2024,Kwok2024}. Moreover, our violent merger explosion can naturally explain most of the other peculiar characteristic properties of this subclass. 

In this picture, the ``bumps'' in the early light curve and the unburned material in early spectra \citep{Dimitriadis2022,Dimitriadis2023,Srivastav2023,Hoogendam2024} are a direct consequence of the circumstellar material that was ejected during the last day before the explosion (see Section~\ref{sec:csm}). This material is absent in double or quadruple detonations \citep{Pakmor2022}, because they already explode before they reach the phase of very high mass transfer rates and bursts of helium burning on the surface of the primary white dwarf that gives rise to it.

The partial burning of the secondary white dwarf naturally explains the presence of unburned material and neon close to the centre of the ejecta (see Figure~\ref{fig:ejecta}), that is required to explain at least some nebular spectra of 03fg-like objects \citep{Taubenberger2019, Dimitriadis2023,Siebert2024,Kwok2024}.

The low ejecta luminosities in some objects \citep[see, for example][]{Taubenberger2011} can in principle be explained by line of sight effects (see Figure~\ref{fig:ejecta}, for lines of sight close to the initial plane of the binary system). These line of sight effects might also be a reason for the additional luminosity at early times. Another contribution could come from interaction with circumstellar material. However, to be relevant, this would likely require a substantial fraction of the original mass of the secondary white dwarf to have ended up in the circumstellar material \citep{Noebauer2016}. Lastly, a wind from the surviving bound object in the centre of the ejecta that interacts with the inner ejecta when they are still optically thick in the centre could also contribute extra luminosity at early times \citep{Shen2017}. More work is clearly needed to really understand the source of the extra luminosity of 03fg-like objects at early times.

Violent mergers have also been associated with 02es-like objects, a subclass of sub-luminous Type Ia supernovae \citep{Maguire2011,Kromer2013,Srivastav2023b,Hoogendam2024}. Because they are sub-luminous, these objects cannot possibly arise from violent merger explosions with massive primary white dwarfs. Despite very good agreement between synthetic observables of violent mergers of lower mass carbon-oxygen white dwarf binaries with observables of 02es-like objects \citep{Kromer2013} it is not obvious how such systems can avoid exploding via double detonation explosions if the helium shell on the primary white dwarf is included.

We speculate that binary systems of almost equal mass carbon-oxygen white dwarfs might also explode as violent mergers, rather than via a double detonation explosion. In this case the double detonation might be avoided because the primary white dwarf and in particular its helium shell is already distorted before the secondary white dwarf overfills its Roche lobe sufficiently to drive a dense enough accretion stream towards the primary white dwarf to ignite it. Such a scenario would bring the violent merger scenario back to its roots where it was introduced for mergers of almost equal mass white dwarfs \citep{Pakmor2010, Pakmor2011, Kromer2013b}.

In this picture, equal mass violent mergers would also be preceded by a characteristic phase in which the two Roche-lobe filling white dwarfs interact, but do not merge yet. In this phase they possibly burn helium-rich material locally, and likely eject significant amounts of material just prior to the explosion to create an envelope of circumstellar material, as seen for 02es-like objects \citep{Hoogendam2024}. The large ejecta mass of the explosion (significantly above the Chandrasekhar-mass) naturally explains the much broader light curves compared to normal sub-luminous Type Ia supernovae \citep{Maguire2011,Ganeshalingam2012,Kromer2013b,Kromer2016}. Finally, equal mass violent mergers could also naturally explain the presence of oxygen in the centre of the ejecta seen in nebular spectra of at least some 02es-like objects \citep{Kromer2013b}. It remains to be seen if equal mass violent mergers can leave bound remnants behind, or if those are an outcome of massive violent mergers only.

The equal mass violent merger scenario will probably need a primary white dwarf with a mass above $\gtrsim 0.8\,\mathrm{M_\odot}$ to produce an observable supernova \citep{Ruiter2013,Shen2018}. Population synthesis studies indicate that most merging binary systems of two carbon-oxygen white dwarfs have a secondary white dwarf with a mass around $0.7\,\mathrm{M_\odot}$ \citep{Liu2018}. The equal mass violent merger scenario predicts that there exists a continuum in brightness of 02es-like objects extending all the way up to 03fg-like objects. Objects at the bright end of 02es-like Type Ia supernovae will be very rare though, but possibly detectable with LSST.

Both sub-classes that have been associated with violent mergers (03fg-like Type Ia supernovae and 02es-like Type Ia supernovae) account for approximately one percent of the Type Ia supernova rate \citep{Dimitriadis2025}. This rate roughly matches the number of fast ($>2000\,\mathrm{km/s}$) hyper-velocity stars \citep{Shen2018,ElBadry2023,Shen2025}. Our simulation is the first to plausibly explain the two extreme hyper-velocity stars D6-1 and D6-3 \citep{Shen2018} that are not only very fast, but also likely of low mass \citep{Shen2025}. This combination is essentially impossible to reach in double detonation explosions where the orbital velocity at explosion and therefore also the ejection velocity of the surviving star increase with the total mass of the system. Interestingly, D6-2, which only moving with $1000\,\mathrm{km/s}$, is much too slow to be explained by the violent merger scenario, but seems to have a similar (very low) mass \citep{Shen2025}. It remains to be shown if violent mergers can also explain the hotter, likely slightly more massive, but also very fast hyper-velocity stars \citep{ElBadry2023,Bhat2025,Glanz2025}.

The rarity of surviving hypervelocity stars and the small contribution of violent mergers to the total Type Ia supernova rate are also consistent with the absence of surviving bound objects in nearby supernova remnants \citep{Schaefer2012, GonzalezHernandez2012,Kerzendorf2018}. Note that even though violent mergers seems to naturally have circumstellar material at the time of explosion, the amount of circumstellar material is likely too low to hide a bound object in a young remnant.

If 03fg-like Type Ia supernovae are indeed violent mergers with massive ($M_\mathrm{primary}\gtrsim1.1\,\mathrm{M_\odot}$) primary white dwarfs, they are almost certainly the tail of primary white dwarfs with the highest masses of a distribution of merging carbon-oxygen white dwarf binaries \citep{Ruiter2013,Liu2018}. Since they contribute $\approx1\%$ to the total Type Ia supernova rate, there will be a large population of merging double white dwarf systems with primary white dwarfs of ($M_\mathrm{primary}<1.1\,\mathrm{M_\odot}$. Most of these primaries will have a more massive helium shell, likely enough to support ignition via the double detonation scenario \citep{Shen2024}. Explosions from these systems will be at least $10$ times more frequent than violent mergers, because mergers of double white dwarf systems with lower mass primary white dwarfs are much more common \citep{Liu2018} and still produce a sufficient amount of radioactive nickel to produce a type Ia supernova \citep{Ruiter2013}. This large population of type Ia supernovae can, by number, only really be explained with normal Type Ia supernovae. They likely explode by igniting the primary white dwarf via double detonation. Usually then both white dwarfs will have to explode in the quadruple detonation scenario \citep{Tanikawa2019,Pakmor2022}, because only a few per cent of the Type Ia supernovae leave a surviving star behind \citep{Shen2025}. More simulations and detailed modelling of quadruple explosions and their observables is needed to understand if these quadruple detonations can fully explain normal Type Ia supernovae \citep{Pakmor2022,Pollin2024,Boos2025,Pollin2025}.

\section{Summary and Outlook}
\label{sec:summary}

We presented and analysed a simulation of a binary system of two carbon-oxygen white dwarfs with the moving-mesh code \textsc{arepo} that follows the inspiral and explosion via the violent merger scenario. In contrast to previous simulations the improved numerical resolution and more accurate treatment of the detonation leaves a bound low-mass ($0.16\,\mathrm{M_\odot}$), fast moving ($2800\,\mathrm{km/s}$) star behind.

We discussed the inspiral, the explosion, and the ejecta structure in Section~\ref{sec:explosion} and found them to be very similar to previous simulations of violent mergers. In particular the ejecta are still highly asymmetric and contain carbon, oxygen, and neon close to the centre. We then looked in detail at the properties of the bound remnant in Section~\ref{sec:remnant} and argued that it is likely consistent with the observed extreme hyper-velocity stars D6-1 and D6-3 and can explain their large velocities as well as the low mass inferred for them.

In Section~\ref{sec:csm} we analysed the material ejected prior to the explosion that creates a circumstellar envelope. We showed that it contains a mass of at least $5\times10^{-3}\mathrm{M_\odot}$, mostly unburned material from the outer layers of the secondary white dwarf, with typical velocities of $1000\,\mathrm{km/s}$ and up to $5000\,\mathrm{km/s}$. This ejected mass is almost certainly only a lower limit, because of the accelerated inspiral we employ and the complete neglect of the earlier phase of mass transfer at much lower mass transfer rates that precedes the phase of the evolution of the binary system when we start our simulation.

Most importantly, we discussed in Section~\ref{sec:discussion} that violent mergers are not only the most likely (and so far only plausible) explanation for the origin of the hyper-velocity stars D6-1 and D6-3, but that they also explain at least qualitatively most characteristic properties of the Type Ia supernova sub-class of "super-Chandrasekhar" or 03fg-like objects. We speculated that equal mass binaries of lower-mass carbon-oxygen white dwarfs could also avoid a double detonation explosion and rather explode as violent mergers. These might explain sub-luminous 02es-like objects. We then argued that if the association of violent mergers with 03fg-like Type Ia supernovae holds and violent mergers are the result of the tail of the distribution of mergers of carbon-oxygen white dwarf binaries with the most massive primary white dwarfs, there has to be a much larger population of mergers with (at least slightly) lower-mass primary white dwarfs that can essentially only lead to normal Type Ia supernovae. 

There are several straightforward next steps to test and confirm our implications. We first need to properly model the long-term evolution of the surviving white dwarf to understand how well it really matches D6-1 and D6-3. We will do so in a companion paper (Bhat et al., in prep). Moreover, we need to run detailed 3D radiative transfer simulations on the ejecta, ideally including polarisation, to understand if the observables of our new simulations still match previous results on early time observables \citep{Pakmor2012,Bulla2016} and late time observables \citep{Blondin2023,Kwok2024}. We also need to better model and understand the impact of circumstellar material from pre-explosion mass loss on observables and confirm that they can explain the early 'bumps' observed for 03fg-like and 02es-like objects \citep{Hoogendam2024}. We need to test if equal mass mergers of lower mass carbon-oxygen white dwarfs lead to violent mergers even if their helium shells are included. We then need to determine more precisely the specific conditions that give rise to a violent merger rather than a double detonation explosion of the primary (and possibly secondary) white dwarf. Finally, we need to better study quadruple detonations to confirm that they are fully consistent with normal Type Ia supernovae.

\begin{acknowledgements}
The authors gratefully acknowledge the Gauss Centre for Supercomputing e.V. (www.gauss-centre.eu) for funding this project by providing computing time on the GCS Supercomputer SuperMUC-NG at Leibniz Supercomputing Centre (www.lrz.de) via the project pn76fu. KJS is supported by NASA through the Astrophysics Theory Program (80NSSC20K0544) and by NASA/ESA Hubble Space Telescope program No.\ 17441. KM acknowledges funding from Horizon Europe ERC grant no. 101125877. FPC would like to acknowledge support from the UK Science and Technology Facilities Council (STFC, grant number ST/X00094X/1). CEC is funded by the European Union’s Horizon Europe research and innovation programme under the Marie Skłodowska-Curie grant agreement No. 101152610. LAK is supported by NASA through a Hubble Fellowship grant No. HF2-51579.001-A awarded by the Space Telescope Science Institute (STScI), which is operated by the Association of Universities for Research in Astronomy, Inc., for NASA, under contract NAS5-26555. Work by EB was performed under the auspices of the U.S. Department of Energy by Lawrence Livermore National Laboratory under Contract DE-AC52-07NA27344. A.B. was supported by the Deutsche Forschungsgemeinschaft (DFG) through grant GE2506/18-1. The work of FKR is supported by the Klaus Tschira Foundation, by the Deutsche Forschungsgemeinschaft (DFG, German Research Foundation) -- RO 3676/7-1, project number 537700965,
and by the European Union (ERC, ExCEED, project number 101096243). Views and opinions expressed are, however, those of the authors only and do not necessarily reflect those of the European Union or the European Research Council Executive Agency. Neither the European Union nor the granting authority can be held responsible for them.
\end{acknowledgements}

\bibliographystyle{aa}

\begin{thebibliography}{73}
\expandafter\ifx\csname natexlab\endcsname\relax\def\natexlab#1{#1}\fi

\bibitem[{{Bauer} {et~al.}(2021){Bauer}, {Chandra}, {Shen}, \&
  {Hermes}}]{Bauer2021}
{Bauer}, E.~B., {Chandra}, V., {Shen}, K.~J., \& {Hermes}, J.~J. 2021, \apjl,
  923, L34

\bibitem[{{Bhat} {et~al.}(2025){Bhat}, {Bauer}, {Pakmor}, {Shen}, {Caiazzo},
  {Rajamuthukumar}, {El-Badry}, \& {Kerzendorf}}]{Bhat2025}
{Bhat}, A., {Bauer}, E.~B., {Pakmor}, R., {et~al.} 2025, \aap, 693, A114

\bibitem[{{Blondin} {et~al.}(2023){Blondin}, {Dessart}, {Hillier},
  {Ramsbottom}, \& {Storey}}]{Blondin2023}
{Blondin}, S., {Dessart}, L., {Hillier}, D.~J., {Ramsbottom}, C.~A., \&
  {Storey}, P.~J. 2023, \aap, 678, A170

\bibitem[{{Boos} {et~al.}(2025){Boos}, {Dessart}, {Townsley}, \&
  {Shen}}]{Boos2025}
{Boos}, S.~J., {Dessart}, L., {Townsley}, D.~M., \& {Shen}, K.~J. 2025, \apj,
  987, 54

\bibitem[{{Braudo} \& {Soker}(2024)}]{Braudo2024}
{Braudo}, J. \& {Soker}, N. 2024, The Open Journal of Astrophysics, 7, 7

\bibitem[{{Bulla} {et~al.}(2016){Bulla}, {Sim}, {Pakmor}, {Kromer},
  {Taubenberger}, {R{\"o}pke}, {Hillebrandt}, \& {Seitenzahl}}]{Bulla2016}
{Bulla}, M., {Sim}, S.~A., {Pakmor}, R., {et~al.} 2016, \mnras, 455, 1060

\bibitem[{{Burmester} {et~al.}(2023){Burmester}, {Ferrario}, {Pakmor},
  {Seitenzahl}, {Ruiter}, \& {Hole}}]{Burmester2023}
{Burmester}, U.~P., {Ferrario}, L., {Pakmor}, R., {et~al.} 2023, \mnras, 523,
  527

\bibitem[{{Cyburt} {et~al.}(2010){Cyburt}, {Amthor}, {Ferguson}, {Meisel},
  {Smith}, {Warren}, {Heger}, {Hoffman}, {Rauscher}, {Sakharuk}, {Schatz},
  {Thielemann}, \& {Wiescher}}]{Cyburt2010}
{Cyburt}, R.~H., {Amthor}, A.~M., {Ferguson}, R., {et~al.} 2010, \apjs, 189,
  240

\bibitem[{{Dimitriadis} {et~al.}(2025){Dimitriadis}, {Burgaz}, {Deckers},
  {Maguire}, {Johansson}, {Smith}, {Rigault}, {Frohmaier}, {Sollerman},
  {Galbany}, {Kim}, {Liu}, {Miller}, {Nugent}, {Alburai}, {Chen}, {Dhawan},
  {Ginolin}, {Goobar}, {Groom}, {Harvey}, {Kenworthy}, {Kulkarni}, {Phan},
  {Popovic}, {Riddle}, {Rusholme}, {M{\"u}ller-Bravo}, {Nordin}, {Terwel}, \&
  {Townsend}}]{Dimitriadis2025}
{Dimitriadis}, G., {Burgaz}, U., {Deckers}, M., {et~al.} 2025, \aap, 694, A10

\bibitem[{{Dimitriadis} {et~al.}(2022){Dimitriadis}, {Foley}, {Arendse},
  {Coulter}, {Jacobson-Gal{\'a}n}, {Siebert}, {Izzo}, {Jones}, {Kilpatrick},
  {Pan}, {Taggart}, {Auchettl}, {Gall}, {Hjorth}, {Kasen}, {Piro}, {Raimundo},
  {Ramirez-Ruiz}, {Rest}, {Swift}, \& {Woosley}}]{Dimitriadis2022}
{Dimitriadis}, G., {Foley}, R.~J., {Arendse}, N., {et~al.} 2022, \apj, 927, 78

\bibitem[{{Dimitriadis} {et~al.}(2023){Dimitriadis}, {Maguire}, {Karambelkar},
  {Lebron}, {Liu}, {Kozyreva}, {Miller}, {Ridden-Harper}, {Anderson}, {Chen},
  {Coughlin}, {Della Valle}, {Drake}, {Galbany}, {Gromadzki}, {Groom},
  {Guti{\'e}rrez}, {Ihanec}, {Inserra}, {Johansson}, {M{\"u}ller-Bravo},
  {Nicholl}, {Polin}, {Rusholme}, {Schulze}, {Sollerman}, {Srivastav},
  {Taggart}, {Wang}, {Yang}, \& {Young}}]{Dimitriadis2023}
{Dimitriadis}, G., {Maguire}, K., {Karambelkar}, V.~R., {et~al.} 2023, \mnras,
  521, 1162

\bibitem[{{Dunkley} {et~al.}(2013){Dunkley}, {Sharpe}, \&
  {Falle}}]{Dunkley2013}
{Dunkley}, S.~D., {Sharpe}, G.~J., \& {Falle}, S. A.~E.~G. 2013, \mnras, 431,
  3429

\bibitem[{{El-Badry} {et~al.}(2023){El-Badry}, {Shen}, {Chandra}, {Bauer},
  {Fuller}, {Strader}, {Chomiuk}, {Naidu}, {Caiazzo}, {Rodriguez}, {Nagarajan},
  {Yamaguchi}, {Vanderbosch}, {Roulston}, {G{\"a}nsicke}, {Han}, {Burdge},
  {Filippenko}, {Brink}, \& {Zheng}}]{ElBadry2023}
{El-Badry}, K., {Shen}, K.~J., {Chandra}, V., {et~al.} 2023, The Open Journal
  of Astrophysics, 6, 28

\bibitem[{{Fryxell} {et~al.}(1989){Fryxell}, {M{\"u}ller}, \&
  {Arnett}}]{Fryxell1989}
{Fryxell}, B., {M{\"u}ller}, E., \& {Arnett}, D. 1989, in Nuclear Astrophysics,
  ed. M.~{Lozano}, M.~I. {Gallardo}, \& J.~M. {Arias}, 100

\bibitem[{{Ganeshalingam} {et~al.}(2012){Ganeshalingam}, {Li}, {Filippenko},
  {Silverman}, {Chornock}, {Foley}, {Matheson}, {Kirshner}, {Milne}, {Calkins},
  \& {Shen}}]{Ganeshalingam2012}
{Ganeshalingam}, M., {Li}, W., {Filippenko}, A.~V., {et~al.} 2012, \apj, 751,
  142

\bibitem[{{Glanz} {et~al.}(2025){Glanz}, {Perets}, {Bhat}, \&
  {Pakmor}}]{Glanz2025}
{Glanz}, H., {Perets}, H.~B., {Bhat}, A., \& {Pakmor}, R. 2025, Nature
  Astronomy

\bibitem[{{Gonz{\'a}lez Hern{\'a}ndez} {et~al.}(2012){Gonz{\'a}lez
  Hern{\'a}ndez}, {Ruiz-Lapuente}, {Tabernero}, {Montes}, {Canal},
  {M{\'e}ndez}, \& {Bedin}}]{GonzalezHernandez2012}
{Gonz{\'a}lez Hern{\'a}ndez}, J.~I., {Ruiz-Lapuente}, P., {Tabernero}, H.~M.,
  {et~al.} 2012, \nat, 489, 533

\bibitem[{{Guillochon} {et~al.}(2010){Guillochon}, {Dan}, {Ramirez-Ruiz}, \&
  {Rosswog}}]{Guillochon2010}
{Guillochon}, J., {Dan}, M., {Ramirez-Ruiz}, E., \& {Rosswog}, S. 2010, \apjl,
  709, L64

\bibitem[{{Hoogendam} {et~al.}(2024){Hoogendam}, {Shappee}, {Brown}, {Tucker},
  {Ashall}, \& {Piro}}]{Hoogendam2024}
{Hoogendam}, W.~B., {Shappee}, B.~J., {Brown}, P.~J., {et~al.} 2024, \apj, 966,
  139

\bibitem[{{Kerzendorf} {et~al.}(2018){Kerzendorf}, {Strampelli}, {Shen},
  {Schwab}, {Pakmor}, {Do}, {Buchner}, \& {Rest}}]{Kerzendorf2018}
{Kerzendorf}, W.~E., {Strampelli}, G., {Shen}, K.~J., {et~al.} 2018, \mnras,
  479, 192

\bibitem[{{Kromer} {et~al.}(2013{\natexlab{a}}){Kromer}, {Fink}, {Stanishev},
  {Taubenberger}, {Ciaraldi-Schoolman}, {Pakmor}, {R{\"o}pke}, {Ruiter},
  {Seitenzahl}, {Sim}, {Blanc}, {Elias-Rosa}, \& {Hillebrandt}}]{Kromer2013}
{Kromer}, M., {Fink}, M., {Stanishev}, V., {et~al.} 2013{\natexlab{a}}, \mnras,
  429, 2287

\bibitem[{{Kromer} {et~al.}(2016){Kromer}, {Fremling}, {Pakmor},
  {Taubenberger}, {Amanullah}, {Cenko}, {Fransson}, {Goobar}, {Leloudas},
  {Taddia}, {R{\"o}pke}, {Seitenzahl}, {Sim}, \& {Sollerman}}]{Kromer2016}
{Kromer}, M., {Fremling}, C., {Pakmor}, R., {et~al.} 2016, \mnras, 459, 4428

\bibitem[{{Kromer} {et~al.}(2013{\natexlab{b}}){Kromer}, {Pakmor},
  {Taubenberger}, {Pignata}, {Fink}, {R{\"o}pke}, {Seitenzahl}, {Sim}, \&
  {Hillebrandt}}]{Kromer2013b}
{Kromer}, M., {Pakmor}, R., {Taubenberger}, S., {et~al.} 2013{\natexlab{b}},
  \apjl, 778, L18

\bibitem[{{Kushnir} \& {Katz}(2020)}]{Kushnir2020b}
{Kushnir}, D. \& {Katz}, B. 2020, \mnras, 493, 5413

\bibitem[{{Kwok} {et~al.}(2024){Kwok}, {Siebert}, {Johansson}, {Jha},
  {Blondin}, {Dessart}, {Foley}, {Hillier}, {Larison}, {Pakmor}, {Temim},
  {Andrews}, {Auchettl}, {Badenes}, {Barnabas}, {Bostroem}, {Brenner Newman},
  {Brink}, {Bustamante-Rosell}, {Camacho-Neves}, {Clocchiatti}, {Coulter},
  {Davis}, {Deckers}, {Dimitriadis}, {Dong}, {Farah}, {Filippenko},
  {Fl{\"o}rs}, {Fox}, {Garnavich}, {Padilla Gonzalez}, {Graur}, {Hambsch},
  {Hosseinzadeh}, {Howell}, {Hughes}, {Kerzendorf}, {Saux}, {Maeda}, {Maguire},
  {McCully}, {Mihalenko}, {Newsome}, {O'Brien}, {Pearson}, {Pellegrino},
  {Pierel}, {Polin}, {Rest}, {Rojas-Bravo}, {Sand}, {Schwab}, {Shahbandeh},
  {Shrestha}, {Smith}, {Strolger}, {Szalai}, {Taggart}, {Terreran}, {Terwel},
  {Tinyanont}, {Valenti}, {Vink{\'o}}, {Wheeler}, {Yang}, {Zheng}, {Ashall},
  {DerKacy}, {Galbany}, {Hoeflich}, {de Jaeger}, {Lu}, {Maund}, {Medler},
  {Morell}, {Shappee}, {Stritzinger}, {Suntzeff}, {Tucker}, \&
  {Wang}}]{Kwok2024}
{Kwok}, L.~A., {Siebert}, M.~R., {Johansson}, J., {et~al.} 2024, \apj, 966, 135

\bibitem[{{Liu} {et~al.}(2018){Liu}, {Wang}, \& {Han}}]{Liu2018}
{Liu}, D., {Wang}, B., \& {Han}, Z. 2018, \mnras, 473, 5352

\bibitem[{{Liu} {et~al.}(2025){Liu}, {Wang}, {Yang}, {Filippenko}, {Brink},
  {Zheng}, {Zhang}, {Li}, \& {Yan}}]{Liu2025}
{Liu}, J., {Wang}, X., {Yang}, Y., {et~al.} 2025, \apjl, 982, L18

\bibitem[{{Liu} {et~al.}(2023){Liu}, {R{\"o}pke}, \& {Han}}]{Liu2023}
{Liu}, Z.-W., {R{\"o}pke}, F.~K., \& {Han}, Z. 2023, Research in Astronomy and
  Astrophysics, 23, 082001

\bibitem[{{Maguire} {et~al.}(2011){Maguire}, {Sullivan}, {Thomas}, {Nugent},
  {Howell}, {Gal-Yam}, {Arcavi}, {Ben-Ami}, {Blake}, {Botyanszki}, {Buton},
  {Cooke}, {Ellis}, {Hook}, {Kasliwal}, {Pan}, {Pereira}, {Podsiadlowski},
  {Sternberg}, {Suzuki}, {Xu}, {Yaron}, {Bloom}, {Cenko}, {Kulkarni}, {Law},
  {Ofek}, {Poznanski}, \& {Quimby}}]{Maguire2011}
{Maguire}, K., {Sullivan}, M., {Thomas}, R.~C., {et~al.} 2011, \mnras, 418, 747

\bibitem[{{Noebauer} {et~al.}(2016){Noebauer}, {Taubenberger}, {Blinnikov},
  {Sorokina}, \& {Hillebrandt}}]{Noebauer2016}
{Noebauer}, U.~M., {Taubenberger}, S., {Blinnikov}, S., {Sorokina}, E., \&
  {Hillebrandt}, W. 2016, \mnras, 463, 2972

\bibitem[{{Ohlmann} {et~al.}(2017){Ohlmann}, {R{\"o}pke}, {Pakmor}, \&
  {Springel}}]{Ohlmann2017}
{Ohlmann}, S.~T., {R{\"o}pke}, F.~K., {Pakmor}, R., \& {Springel}, V. 2017,
  \aap, 599, A5

\bibitem[{{Pakmor} {et~al.}(2022){Pakmor}, {Callan}, {Collins}, {de Mink},
  {Holas}, {Kerzendorf}, {Kromer}, {Neunteufel}, {O'Brien}, {R{\"o}pke},
  {Ruiter}, {Seitenzahl}, {Shingles}, {Sim}, \& {Taubenberger}}]{Pakmor2022}
{Pakmor}, R., {Callan}, F.~P., {Collins}, C.~E., {et~al.} 2022, \mnras, 517,
  5260

\bibitem[{{Pakmor} {et~al.}(2012{\natexlab{a}}){Pakmor}, {Edelmann},
  {R{\"o}pke}, \& {Hillebrandt}}]{Pakmor2012}
{Pakmor}, R., {Edelmann}, P., {R{\"o}pke}, F.~K., \& {Hillebrandt}, W.
  2012{\natexlab{a}}, \mnras, 424, 2222

\bibitem[{{Pakmor} {et~al.}(2011){Pakmor}, {Hachinger}, {R{\"o}pke}, \&
  {Hillebrand t}}]{Pakmor2011}
{Pakmor}, R., {Hachinger}, S., {R{\"o}pke}, F.~K., \& {Hillebrand t}, W. 2011,
  \aap, 528, A117

\bibitem[{{Pakmor} {et~al.}(2010){Pakmor}, {Kromer}, {R{\"o}pke}, {Sim},
  {Ruiter}, \& {Hillebrandt}}]{Pakmor2010}
{Pakmor}, R., {Kromer}, M., {R{\"o}pke}, F.~K., {et~al.} 2010, \nat, 463, 61

\bibitem[{{Pakmor} {et~al.}(2012{\natexlab{b}}){Pakmor}, {Kromer},
  {Taubenberger}, {Sim}, {R{\"o}pke}, \& {Hillebrandt}}]{Pakmor2012b}
{Pakmor}, R., {Kromer}, M., {Taubenberger}, S., {et~al.} 2012{\natexlab{b}},
  \apjl, 747, L10

\bibitem[{{Pakmor} {et~al.}(2013){Pakmor}, {Kromer}, {Taubenberger}, \&
  {Springel}}]{Pakmor2013}
{Pakmor}, R., {Kromer}, M., {Taubenberger}, S., \& {Springel}, V. 2013, \apjl,
  770, L8

\bibitem[{{Pakmor} {et~al.}(2016){Pakmor}, {Springel}, {Bauer}, {Mocz},
  {Munoz}, {Ohlmann}, {Schaal}, \& {Zhu}}]{Pakmor2016}
{Pakmor}, R., {Springel}, V., {Bauer}, A., {et~al.} 2016, \mnras, 455, 1134

\bibitem[{{Pakmor} {et~al.}(2021){Pakmor}, {Zenati}, {Perets}, \&
  {Toonen}}]{Pakmor2021}
{Pakmor}, R., {Zenati}, Y., {Perets}, H.~B., \& {Toonen}, S. 2021, \mnras, 503,
  4734

\bibitem[{{Paxton} {et~al.}(2011){Paxton}, {Bildsten}, {Dotter}, {Herwig},
  {Lesaffre}, \& {Timmes}}]{Paxton2011}
{Paxton}, B., {Bildsten}, L., {Dotter}, A., {et~al.} 2011, \apjs, 192, 3

\bibitem[{{Pollin} {et~al.}(2024){Pollin}, {Sim}, {Pakmor}, {Callan},
  {Collins}, {Shingles}, {R{\"o}pke}, \& {Srivastav}}]{Pollin2024}
{Pollin}, J.~M., {Sim}, S.~A., {Pakmor}, R., {et~al.} 2024, \mnras, 533, 3036

\bibitem[{{Pollin} {et~al.}(2025){Pollin}, {Sim}, {Shingles}, {Pakmor},
  {Callan}, {Collins}, {Roepke}, {Kwok}, {Holas}, \& {Srivastav}}]{Pollin2025}
{Pollin}, J.~M., {Sim}, S.~A., {Shingles}, L.~J., {et~al.} 2025, arXiv
  e-prints, arXiv:2507.05000, subm. MNRAS

\bibitem[{{Raskin} \& {Kasen}(2013)}]{Raskin2013}
{Raskin}, C. \& {Kasen}, D. 2013, \apj, 772, 1

\bibitem[{{Ruiter} \& {Seitenzahl}(2025)}]{Ruiter2025}
{Ruiter}, A.~J. \& {Seitenzahl}, I.~R. 2025, \aapr, 33, 1

\bibitem[{{Ruiter} {et~al.}(2013){Ruiter}, {Sim}, {Pakmor}, {Kromer},
  {Seitenzahl}, {Belczynski}, {Fink}, {Herzog}, {Hillebrandt}, {R{\"o}pke}, \&
  {Taubenberger}}]{Ruiter2013}
{Ruiter}, A.~J., {Sim}, S.~A., {Pakmor}, R., {et~al.} 2013, \mnras, 429, 1425

\bibitem[{{Saio} \& {Nomoto}(1985)}]{Saio1985}
{Saio}, H. \& {Nomoto}, K. 1985, \aap, 150, L21

\bibitem[{{Sato} {et~al.}(2015){Sato}, {Nakasato}, {Tanikawa}, {Nomoto},
  {Maeda}, \& {Hachisu}}]{Sato2015}
{Sato}, Y., {Nakasato}, N., {Tanikawa}, A., {et~al.} 2015, \apj, 807, 105

\bibitem[{{Sato} {et~al.}(2016){Sato}, {Nakasato}, {Tanikawa}, {Nomoto},
  {Maeda}, \& {Hachisu}}]{Sato2016}
{Sato}, Y., {Nakasato}, N., {Tanikawa}, A., {et~al.} 2016, \apj, 821, 67

\bibitem[{{Schaefer} \& {Pagnotta}(2012)}]{Schaefer2012}
{Schaefer}, B.~E. \& {Pagnotta}, A. 2012, \nat, 481, 164

\bibitem[{{Schwab}(2021)}]{Schwab2021}
{Schwab}, J. 2021, \apj, 906, 53

\bibitem[{{Schwab} {et~al.}(2012){Schwab}, {Shen}, {Quataert}, {Dan}, \&
  {Rosswog}}]{Schwab2012}
{Schwab}, J., {Shen}, K.~J., {Quataert}, E., {Dan}, M., \& {Rosswog}, S. 2012,
  \mnras, 427, 190

\bibitem[{{Seitenzahl} {et~al.}(2009){Seitenzahl}, {Meakin}, {Townsley},
  {Lamb}, \& {Truran}}]{Seitenzahl2009}
{Seitenzahl}, I.~R., {Meakin}, C.~A., {Townsley}, D.~M., {Lamb}, D.~Q., \&
  {Truran}, J.~W. 2009, \apj, 696, 515

\bibitem[{{Shen}(2025)}]{Shen2025}
{Shen}, K.~J. 2025, \apj, 982, 6

\bibitem[{{Shen} {et~al.}(2012){Shen}, {Bildsten}, {Kasen}, \&
  {Quataert}}]{Shen2012}
{Shen}, K.~J., {Bildsten}, L., {Kasen}, D., \& {Quataert}, E. 2012, \apj, 748,
  35

\bibitem[{{Shen} {et~al.}(2023){Shen}, {Blouin}, \& {Breivik}}]{Shen2023}
{Shen}, K.~J., {Blouin}, S., \& {Breivik}, K. 2023, \apjl, 955, L33

\bibitem[{{Shen} {et~al.}(2024){Shen}, {Boos}, \& {Townsley}}]{Shen2024}
{Shen}, K.~J., {Boos}, S.~J., \& {Townsley}, D.~M. 2024, \apj, 975, 127

\bibitem[{{Shen} {et~al.}(2018){Shen}, {Kasen}, {Miles}, \&
  {Townsley}}]{Shen2018}
{Shen}, K.~J., {Kasen}, D., {Miles}, B.~J., \& {Townsley}, D.~M. 2018, \apj,
  854, 52

\bibitem[{{Shen} \& {Moore}(2014)}]{Shen2014b}
{Shen}, K.~J. \& {Moore}, K. 2014, \apj, 797, 46

\bibitem[{{Shen} \& {Schwab}(2017)}]{Shen2017}
{Shen}, K.~J. \& {Schwab}, J. 2017, \apj, 834, 180

\bibitem[{{Siebert} {et~al.}(2023){Siebert}, {Foley}, {Zenati}, {Dimitriadis},
  {Schmidt}, {Yang}, {Davis}, {Taggart}, \& {Rojas-Bravo}}]{Siebert2023}
{Siebert}, M.~R., {Foley}, R.~J., {Zenati}, Y., {et~al.} 2023, \apj, 958, 173

\bibitem[{Siebert {et~al.}(2024)Siebert, Kwok, Johansson, Jha, Blondin,
  Dessart, Foley, Hillier, Larison, Pakmor, Temim, Andrews, Auchettl, Badenes,
  Barna, Bostroem, Newman, Brink, Bustamante-Rosell, Camacho-Neves,
  Clocchiatti, Coulter, Davis, Deckers, Dimitriadis, Dong, Farah, Filippenko,
  Flörs, Fox, Garnavich, Gonzalez, Graur, Hambsch, Hosseinzadeh, Howell,
  Hughes, Kerzendorf, Saux, Maeda, Maguire, McCully, Mihalenko, Newsome,
  O’Brien, Pearson, Pellegrino, Pierel, Polin, Rest, Rojas-Bravo, Sand,
  Schwab, Shahbandeh, Shrestha, Smith, Strolger, Szalai, Taggart, Terreran,
  Terwel, Tinyanont, Valenti, Vinkó, Wheeler, Yang, Zheng, Ashall, DerKacy,
  Galbany, Hoeflich, Hsiao, de~Jaeger, Lu, Maund, Medler, Morrell, Shappee,
  Stritzinger, Suntzeff, Tucker, \& Wang}]{Siebert2024}
Siebert, M.~R., Kwok, L.~A., Johansson, J., {et~al.} 2024, The Astrophysical
  Journal, 960, 88

\bibitem[{{Springel}(2005)}]{Springel2005}
{Springel}, V. 2005, \mnras, 364, 1105

\bibitem[{{Springel}(2010)}]{Arepo}
{Springel}, V. 2010, \mnras, 401, 791

\bibitem[{{Srivastav} {et~al.}(2023{\natexlab{a}}){Srivastav}, {Moore},
  {Nicholl}, {Magee}, {Smartt}, {Fulton}, {Sim}, {Pollin}, {Galbany},
  {Inserra}, {Kozyreva}, {Moriya}, {Callan}, {Sheng}, {Smith}, {Sommer},
  {Anderson}, {Deckers}, {Gromadzki}, {M{\"u}ller-Bravo}, {Pignata}, {Rest}, \&
  {Young}}]{Srivastav2023b}
{Srivastav}, S., {Moore}, T., {Nicholl}, M., {et~al.} 2023{\natexlab{a}},
  \apjl, 956, L34

\bibitem[{{Srivastav} {et~al.}(2023{\natexlab{b}}){Srivastav}, {Smartt},
  {Huber}, {Dimitriadis}, {Chambers}, {Fulton}, {Moore}, {Callan},
  {Gillanders}, {Maguire}, {Nicholl}, {Shingles}, {Sim}, {Smith}, {Anderson},
  {de Boer}, {Chen}, {Gao}, \& {Young}}]{Srivastav2023}
{Srivastav}, S., {Smartt}, S.~J., {Huber}, M.~E., {et~al.} 2023{\natexlab{b}},
  \apjl, 943, L20

\bibitem[{{Tanikawa} {et~al.}(2019){Tanikawa}, {Nomoto}, {Nakasato}, \&
  {Maeda}}]{Tanikawa2019}
{Tanikawa}, A., {Nomoto}, K., {Nakasato}, N., \& {Maeda}, K. 2019, \apj, 885,
  103

\bibitem[{{Taubenberger}(2017)}]{Taubenberger2017}
{Taubenberger}, S. 2017, in Handbook of Supernovae, ed. A.~W. {Alsabti} \&
  P.~{Murdin}, 317

\bibitem[{{Taubenberger} {et~al.}(2011){Taubenberger}, {Benetti}, {Childress},
  {Pakmor}, {Hachinger}, {Mazzali}, {Stanishev}, {Elias-Rosa}, {Agnoletto},
  {Bufano}, {Ergon}, {Harutyunyan}, {Inserra}, {Kankare}, {Kromer},
  {Navasardyan}, {Nicolas}, {Pastorello}, {Prosperi}, {Salgado}, {Sollerman},
  {Stritzinger}, {Turatto}, {Valenti}, \& {Hillebrandt}}]{Taubenberger2011}
{Taubenberger}, S., {Benetti}, S., {Childress}, M., {et~al.} 2011, \mnras, 412,
  2735

\bibitem[{{Taubenberger} {et~al.}(2019){Taubenberger}, {Floers}, {Vogl},
  {Kromer}, {Spyromilio}, {Aldering}, {Antilogus}, {Bailey}, {Baltay},
  {Bongard}, {Boone}, {Buton}, {Chotard}, {Copin}, {Dixon}, {Fouchez},
  {Fransson}, {Gangler}, {Gupta}, {Hachinger}, {Hayden}, {Hillebrandt}, {Kim},
  {Kowalski}, {Leget}, {Leibundgut}, {Mazzali}, {Noebauer}, {Nordin}, {Pain},
  {Pakmor}, {Pecontal}, {Pereira}, {Perlmutter}, {Ponder}, {Rabinowitz},
  {Rigault}, {Rubin}, {Runge}, {Saunders}, {Smadja}, {Tao}, \&
  {Thomas}}]{Taubenberger2019}
{Taubenberger}, S., {Floers}, A., {Vogl}, C., {et~al.} 2019, \mnras, 488, 5473

\bibitem[{{Taubenberger} {et~al.}(2013{\natexlab{a}}){Taubenberger}, {Kromer},
  {Hachinger}, {Mazzali}, {Benetti}, {Nugent}, {Scalzo}, {Pakmor}, {Stanishev},
  {Spyromilio}, {Bufano}, {Sim}, {Leibundgut}, \&
  {Hillebrandt}}]{Taubenberger2013b}
{Taubenberger}, S., {Kromer}, M., {Hachinger}, S., {et~al.} 2013{\natexlab{a}},
  \mnras, 432, 3117

\bibitem[{{Taubenberger} {et~al.}(2013{\natexlab{b}}){Taubenberger}, {Kromer},
  {Pakmor}, {Pignata}, {Maeda}, {Hachinger}, {Leibundgut}, \&
  {Hillebrandt}}]{Taubenberger2013}
{Taubenberger}, S., {Kromer}, M., {Pakmor}, R., {et~al.} 2013{\natexlab{b}},
  \apjl, 775, L43

\bibitem[{{Timmes} \& {Swesty}(2000)}]{Timmes2000}
{Timmes}, F.~X. \& {Swesty}, F.~D. 2000, \apjs, 126, 501

\bibitem[{{Weinberger} {et~al.}(2020){Weinberger}, {Springel}, \&
  {Pakmor}}]{Weinberger2020}
{Weinberger}, R., {Springel}, V., \& {Pakmor}, R. 2020, \apjs, 248, 32

\end{thebibliography}

\end{document}